\documentclass{PoS}
\pdfoutput=1

\graphicspath{ {figs/} }

\title{$B_s\to \mu^+\mu^-$ and $\bar{B} \to X_s \gamma$ to NNLO}

\ShortTitle{$B_s\to \mu^+\mu^-$ and $\bar{B} \to X_s \gamma$ to NNLO}

\author{\speaker{Matthias Steinhauser}\thanks{This work was supported by the
    DFG through the SFB/TR~9 ``Computational Particle Physics''.
    I would like to thank 
    Christoph Bobeth, Martin Gorbahn, Thomas Hermann, Miko{\l}aj
    Misiak, and Emmanuel Stamou for a fruitful collaboration on the subject of
    this contribution.}\\
        Institut f\"ur Theoretische Teilchenphysik, Karlsruhe
        Institute of Technology (KIT),\\ D-76128 Karlsruhe, Germany\\
        E-mail: \email{matthias.steinhauser@kit.edu}}

\abstract{In this contribution, the recent calculation of
  next-to-next-to-leading order QCD and next-to-leading order electroweak
  corrections to the decay $B_s\to \mu^+\mu^-$ is reviewed, and a
  detailed discussion of the uncertainty of the theory prediction is
  provided. Furthermore, we discuss the status of the next-to-next-to-leading 
  order QCD corrections to $\bar{B} \to X_s \gamma$.}

\FullConference{ Loops and Legs in Quantum Field Theory - LL 2014,\\
		 27 April - 2 May 2014 \\
		 Weimar, Germany }

\begin{document}

\section{Introduction}

In the last decades, the Standard Model (SM) of particle physics has been
confirmed at many different energy scales, and there are only a few processes
where deviations at the level of $2-3\sigma$ are observed. Nevertheless, there
are several questions where the SM cannot provide a satisfactory answer, which
triggers a huge activity --- both on the theory and on the experimental side
--- to search for deviations and thus for hints of physical phenomena in
accordance with extended theories.  A very promising place to look for
discrepancies with SM predictions are processes related to $B$ mesons since on
the one hand there are precise measurements available, and on the other hand it
is possible to perform precision calculations, and to arrive at rigorous
predictions. Among the most promising processes there are the decays of a
$B_s$ meson into two muons, and of a $B$ meson into a meson containing a
strange quark and a photon.  It is common to both of them that the SM
contribution is loop-induced, and thus potential new physics effects are
parametrically of the same order of magnitude, which can hence lead to sizable
effects. Whereas the branching ratio in the case of $\bar{B} \to X_s \gamma$
is of the order $10^{-4}$, it is of the order $10^{-9}$ for $B_s\to\mu^+\mu^-$.

The main part of this proceedings contribution deals with $B_s\to \mu^+\mu^-$.
End of 2013 next-to-next-to-leading order (NNLO) QCD and NLO electroweak (EW)
corrections became
available~\cite{Hermann:2013kca,Bobeth:2013tba,Bobeth:2013uxa} which we
briefly review in Section~\ref{sec::calc}. Afterwards, in
Section~\ref{sec::phen}, the theory prediction is discussed where special
emphasis is put on the corresponding uncertainty.

NNLO corrections to $\bar{B} \to X_s \gamma$
have been computed in Refs.~\cite{Misiak:2006zs,Misiak:2006ab}. 
In Section~\ref{sec::bsg} we provide a brief status of the 
updated prediction for the branching ratio based
on improved theoretical input.

\section{\label{sec::calc}NNLO QCD and NLO EW corrections to $B_s\to \mu^+\mu^-$}

There are several energy scales involved in decays of $B$ mesons: first of all
there is the typical scale of the decay process, $\mu_b$, which is of the
order of the bottom quark mass. Furthermore, there are masses of the
virtual particles in the loops, which in the SM are essentially given by the
$W$ boson and the top quark.  In the following, the latter scale is denoted by
$\mu_0$.  The framework which can be used to perform calculations involving
widely separated scales is based on an effective theory where the heavy degrees
of freedom are integrated out from the underlying theory.  In the
case at hand, this leads to an effective Lagrange density with 
only one relevant effective operator\footnote{In beyond-SM theories sizable
  contributions are also obtained from operators with scalar and pseudo-scalar
  currents.}
$Q_A = (\bar{b} \gamma_{\alpha} \gamma_5 s)(\bar{\mu} \gamma^{\alpha}\gamma_5 \mu)$.
In order to arrive at predictions for the decay rate, the corresponding
matching coefficient, $C_A$, has to be computed in a first step at the 
high scale $\mu_0$. Afterwards, it has to be evolved to $\mu_b$ using
renormalization group techniques. 

The one-loop calculation of $C_A$ has been performed for the first time in
Ref.~\cite{Inami:1980fz}, and NLO QCD corrections have been considered
in Refs.~\cite{Buchalla:1992zm,Buchalla:1993bv,Misiak:1999yg,Buchalla:1998ba}.
Recently, the three-loop corrections, $C_A^{(2)}$, have been computed in
Ref.~\cite{Hermann:2013kca}. The calculation can be reduced to vacuum
integrals involving the two masses $M_W$ and $m_t$, which have been solved with
the help of expansions for $m_t\gg M_W$ and $m_t\approx M_W$. In this way,
only one-scale integrals have to be computed and simple analytic results
are obtained, which are of polynomial form with at most logarithmic
coefficients. Thus, the numerical evaluation is simple and straightforward.

\begin{figure}[t]
  \begin{center}
    \includegraphics[width=.6\textwidth]{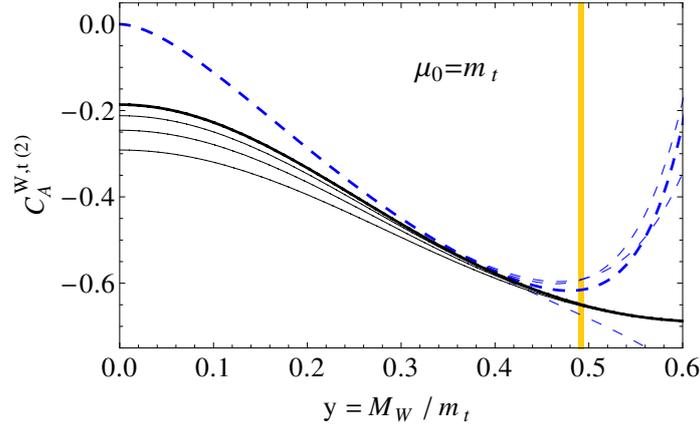}
    \caption{\label{fig::bs_CA}Three-loop corrections to the $W$-box
      contribution to $C_A^{(2)}$ involving 
      $W$-top-bottom vertices. Solid lines result from
      expansions around $y=1$ and dashed lines are obtained for $y\ll 1$.}
  \end{center}
\end{figure}

As an exemplary case we show in Fig.~\ref{fig::bs_CA} the result for the
$W$-box contribution to $C_A^{(2)}$ from diagrams with $W$-top-bottom vertices
as a function of $y=M_W/m_t$. The solid (black) lines originate from the
expansion around $y=1$, and the dashed (blue) lines from the case
$y\ll1$. Thinner lines correspond to results involving less expansion
terms. One observes that deviations only occur either below or above $y\approx
0.4$, which leads to the conclusion that the combination of the expansions
provide an excellent approximation to $C_A^{(2)}$ in the whole range
$y\in[0,1]$. In particular, in the physical region for $y$, which is indicated
by the vertical (yellow) band, the expansion around the equal-mass limit alone
is sufficient to provide a three-loop prediction.  The inclusion of the
three-loop QCD corrections to $C_A$ reduces the uncertainties to the branching
ratio from scale variation of $\mu_0$ in the interval
$m_t/2$ to $2 m_t$ from 1.8\% to below 0.2\%~\cite{Hermann:2013kca}.

\begin{figure}[t]
  \begin{center}
    \includegraphics[width=1.0\textwidth]{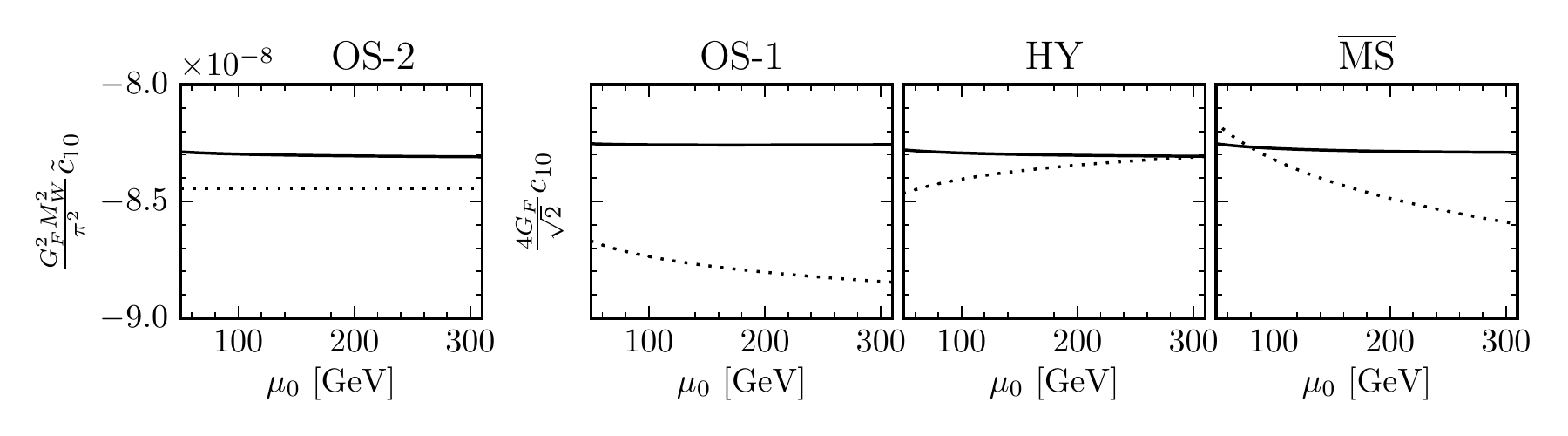}
    \caption{\label{fig::bs_ew}NLO electroweak corrections adopting different
      renormalization schemes.}
  \end{center}
\end{figure}

The complete NLO EW corrections have been computed in
Ref.~\cite{Bobeth:2013tba}. Before, only the leading $m_t^2$ terms were known
from Ref.~\cite{Buchalla:1997kz} and the corresponding uncertainties have been
estimated to be of the order of about 7\%.  In contrast to QCD, there is a
non-trivial running of $C_A$ from $\mu_0$ to $\mu_b$ after including ${\cal
  O}(\alpha_{em})$ terms which originate from mixing of operators once QED
corrections are turned on~\cite{Bobeth:2003at,Huber:2005ig,Misiak:2011bf}.  In
Ref.~\cite{Bobeth:2013tba} this effect has been taken into account together
with a detailed study of the renormalization scheme dependence. The results are
shown in Fig.~\ref{fig::bs_ew} where $\tilde{c}_{10}=-2C_A$ is plotted for
four different choices (see Ref.~\cite{Bobeth:2013tba} for details) at LO and
NLO. One observes huge differences at LO (dotted line) which basically
disappear at NLO (solid line).

\section{\label{sec::phen}Prediction for $\overline{\mathcal B}(B_s\to \mu^+ \mu^-)$}

In 2013 the CMS and LHCb experiments at the LHC have provided 
measurements for the averaged time-integrated branching ratios for $B_s\to
\mu^+ \mu^-$ and $B_d\to \mu^+ \mu^-$. The
combination~\cite{LHCb-CMS-combi:2013} given by
\begin{equation} 
  \label{eq::brexp}
  {\overline{\mathcal B}_{s\mu}} = (2.9 \pm 0.7) \times 10^{-9},\hspace{5mm}
  {\overline{\mathcal B}_{d\mu}} = \left(3.6^{+1.6}_{-1.4}\right) \times
  10^{-10}
  \,,
\end{equation}
is based on the measurements~\cite{Chatrchyan:2013bka}
and~\cite{Aaij:2013aka}. In the forthcoming decade a reduction of the
uncertainties to a few percent level is expected, in particular for
${\overline{\mathcal B}_{s\mu}}$, in which case the uncertainty 
is to a large extent dominated by
statistics. In Ref.~\cite{Bobeth:2013uxa} the numbers in Eq.~(\ref{eq::brexp})
have been confronted with theory predictions including NNLO QCD and NLO EW
corrections, as discussed in Section~\ref{sec::calc}.
In the following, we concentrate on $B_s\to \mu^+ \mu^-$; the arguments hold
analogously also for $B_d\to \mu^+ \mu^-$.

Within the effective-theory framework the branching ratio $\overline{\mathcal
  B}_{s\mu}$ can be cast in the form (see also Ref.~\cite{DeBruyn:2012wk})
\begin{equation} 
  \label{eq::br1}
  \overline{\mathcal B}_{s\mu} 
  = \frac{|N|^2 M_{B_s}^3 f_{B_s}^2}{8\pi\,\Gamma^s_H}\, 
  \beta_{s\mu}\, r_{s\mu}^2\, |C_A(\mu_b)|^2 \,+\, {\mathcal O}(\alpha_{em})
  \,,
\end{equation}
with $N = V_{tb}^\star V_{ts}^{}\, G_F^2 M_W^2/ \pi^2$, $r_{s\mu} = 2
m_\mu/M_{B_s}$, $\beta_{s\mu} = \sqrt{1-r_{s\mu}^2}$ and $\Gamma^s_H$
denoting the heavier mass-eigenstate total width.  $M_{B_s}$ is the
$B_s$-meson mass, and $f_{B_s}$ its decay constant which is defined by the QCD
matrix element $\langle 0| \bar b \gamma^\alpha \gamma_5 s | B_s(p) \rangle =
i p^\alpha f_{B_s}$.
The term ``${\mathcal O}(\alpha_{em})$'' originates from the fact
that NLO QED corrections to $\overline{\mathcal B}_{s\mu}$
are not complete as virtual NLO corrections to the matrix element
$\langle \mu^+\mu^-|Q_A|B_s\rangle$ are missing. 
It has been estimated to be of the order of $0.3\%$,
which is based on the following observations:
\begin{figure}[t]
  \begin{center}
    \includegraphics[width=.6\textwidth]{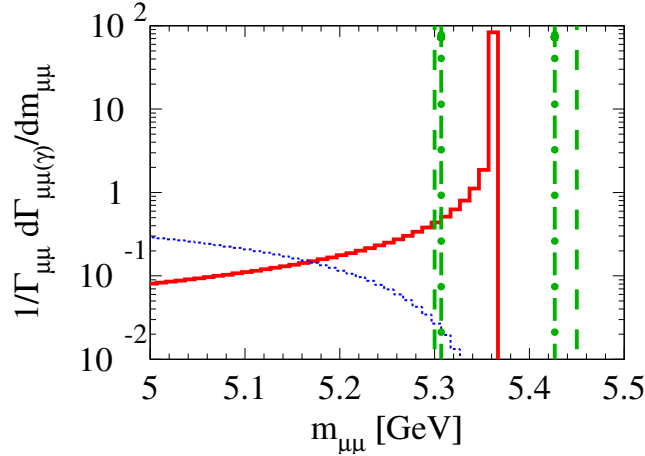}
    \caption{\label{fig::gammumu}Contributions to the dimuon invariant-mass
      spectrum in $B_s \to \mu^+ \mu^- (n\gamma)$ with $n=0,1,2,\ldots$ The
      (blue) dotted curve describes the contribution from photons radiated off
      a quark line, whereas the (red) solid line originates from photonic
      corrections to the muon line.  Both of them are displayed in bins of
      $0.01\,$GeV width.  The (green) vertical lines indicate the blinded
      signal windows of the CMS (dashed) and LHCb (dash-dotted) experiments.}
  \end{center}
\end{figure}
\begin{itemize}
\item The ${\mathcal O}(\alpha_{em})$ term in Eq.~(\ref{eq::br1}) does not 
  contain any enhancement factor like $m_t^2/M_W^2$ or $1/\sin^2\Theta_W$.
  Such factors are present in the genuine NLO EW corrections
  of Ref.~\cite{Bobeth:2013tba}.

\item Soft photon bremsstrahlung can potentially lead to sizeable ${\mathcal
    O}(\alpha_{em})$ corrections.  The dimuon invariant-mass spectrum for the
  decay rate $B_s \to \mu^+ \mu^- (n\gamma)$ with $n=0,1,2,\ldots\,$ is shown
  in Fig.~\ref{fig::gammumu} where the (blue) dotted curve corresponds to real
  photon emission from the quarks, which has been discussed in
  Ref.~\cite{Aditya:2012im}.  The red curve is supposed to contain all other
  ${\mathcal O}(\alpha_{em})$ corrections, in particular the soft photon
  radiation from the muons~\cite{Buras:2012ru} which constitutes the dominant
  part.

  As can be seen, the (blue) dotted curve is highly suppressed in the signal
  regions of CMS and LHCb which are indicated by the dashed and dash-dotted
  vertical lines. This contribution is infrared safe because the decaying
  meson is electrically neutral. On the experimental side it is (should be)
  treated as background and it is not considered at all in the theory
  prediction.

  On the other hand, the photon bremsstrahlung from the muons is taken into
  account in both experimental analyses with the help of the program
  PHOTOS~\cite{Golonka:2005pn} which is used to extra\-polate along the solid
  (red) curve down to zero. Thus, in the resulting quantity there are no large QED
  logarithms, which in principle could emerge from the applied cuts, and one
  remains with a usual ${\mathcal O}(\alpha_{em})$ correction without extra
  enhancement.

\item One has to worry about the question whether it is possible that QED
  corrections of order $\alpha_{em}/\pi \approx 2\times 10^{-3}$ can remove a
  helicity suppression factor $m_\mu^2/M_{B_s}^2 \approx 10^{-4}$ and thus
  lead to correction terms which have to be taken into account. The following
  reasons show that this is not the case
  \begin{itemize}
  \item Virtual ${\cal O}(\alpha_{em})$ corrections both on the quark and
    muon line cannot undo the helicity
    suppression in the SM since then the same argument holds as at LO
    and the vector or axial-vector lepton
    currents will always lead to suppression factors $m_\mu^2/M_{B_s}^2$ 
    in the branching ratio.
  \item Real photon corrections of order $\alpha_{em}$ off a muon 
    still lead to helicity suppression because the QCD matrix element
    defining $f_{B_s}$ is proportional to $p^\alpha$ which is the momentum 
    flowing into the vector or axial-vector lepton current.
    Since the latter is conserved for massless leptons, helicity suppression
    remains in action.
  \item Real photon corrections from the quarks can lift the helicity
    suppression, however, the corresponding contribution is highly phase space
    suppressed in the signal region~\cite{Aditya:2012im}, 
    as is shown by the blue line in
    Fig.~\ref{fig::gammumu}.
  \item Real photon corrections from the muon can lift the helicity
    suppression only in case there is a further virtual photon connecting the
    quark and muon line.  In that case the above argument is not valid,
    however, we have a suppression factor $\alpha_{em}^3$ for the branching
    ratio, which is significantly smaller than $m_\mu^2/M_{B_s}^2$.
  \end{itemize}

\item The ${\mathcal O}(\alpha_{em})$ has to cancel the $\mu_b$ dependence of
  $|C_A(\mu_b)|^2$ which amounts to $0.3\%$ when varying $\mu_b$ between
  $m_b/2$ and $2m_b$. 
\end{itemize}

\begin{table}
\begin{center}
\renewcommand{\arraystretch}{1.4}
\begin{tabular}{c|c|ccc|cc|c|c|c}
& 
& $f_{B_q}$ 
& CKM 
& $\tau_H^q$ 
& $M_t$ 
& $\alpha_s$ 
& other      
& non-        
& $\sum$
\\[-2mm]
& & & & & & & param. & param. &  
\\
\hline
  $\overline{\mathcal{B}}_{s\mu}$
& $(3.65 \pm 0.23) \times 10^{-9}$ 
& $4.0$\%
& $4.3\%$ 
& $1.3$\%
& $1.6$\%
& $0.1$\%
& $< 0.1$\%
& $1.5$\%
& $6.4$\%
\\
  $\overline{\mathcal{B}}_{d\mu}$
& $(1.06 \pm 0.09) \times 10^{-10}$ 
& $4.5$\%
& $6.9$\%
& $0.5$\%
& $1.6$\%
& $0.1$\%
& $< 0.1$\%
& $1.5$\%
& $8.5$\%
\end{tabular}
\renewcommand{\arraystretch}{1.0}
\caption{\label{tab::BRresults} 
  Central value and relative uncertainties from various sources 
  for $\overline{\mathcal{B}}_{s\mu}$ 
  and $\overline{\mathcal{B}}_{d\mu}$. 
  In the last column they are added in quadrature.}
\end{center}
\end{table}

At that point it is straightforward to evaluate the branching ratio. We refrain
from repeating the discussion about the 
input parameters which can be found in Ref.~\cite{Bobeth:2013uxa} 
but want to mention a few important issues in connection to the uncertainty of
the branching ratio. A summary is given in Table~\ref{tab::BRresults}.
\begin{itemize}
\item 
  The largest contributions to the uncertainties arise from
  the decay constants and the CKM matrix elements. The former is based on
  lattice determinations, and the values for $f_{B_s}$ and $f_{B_d}$ are taken
  over from a compilation of the Flavour Lattice Averaging Group
  (FLAG)~\cite{Aoki:2013ldr}.
\item 
  In the case of $B_s$, we write the CKM factors $|V_{tb}^\star V_{ts}^{}|$ as
  $|V_{cb}| \times |V_{tb}^\star V_{ts}^{}/V_{cb}^{}|$, which allows us to use
  numerical results for the accurately known ratio $|V_{tb}^\star
  V_{ts}^{}/V_{cb}^{}|$.  Furthermore, a precise result for $|V_{cb}|$ has
  recently be obtained in Ref.~\cite{Gambino:2013rza} taking into account both
  the semileptonic data and the precise quark mass determinations from
  flavor-conserving processes.
\item
  The parameters $M_t$ (on-shell top quark mass)
  and $\alpha_s$ enter the matching coefficient $C_A(\mu_b)$ in a non-trivial
  way. In Ref.~\cite{Bobeth:2013uxa} formulae for the branching ratio are
  provided which allow for a convenient change of these parameters.
\item 
  The column ``other param.'' in Table~\ref{tab::BRresults}
  shows that the uncertainties originating from the not explicitly listed
  parameters (like the Higgs or gauge boson masses or the Fermi constant)
  are negligible.
\item 
  The contributions to the non-parametric uncertainties, which
  are estimated to 1.5\% both for $\overline{\mathcal{B}}_{s\mu}$ and
  $\overline{\mathcal{B}}_{d\mu}$ include
  \\
  \mbox{}\hspace*{3em}
  \begin{tabular}{lc}
    ${\mathcal O}(\alpha_{em})$ in Eq.~(\ref{eq::br1}) & 0.3\% \\
    NNLO QCD $\mu_0$ dependence                        & 0.2\% \\
    NLO EW $\mu_0$ dependence                          & 0.2\% \\
    NLO EW renormalization scheme dependence           & 0.6\% \\
    Higher-order $M_{B_q}^2/M_W^2$ power corrections   & 0.4\% \\
    $\overline{\rm MS}-$OS top quark mass conversion   & 0.3\% \\
  \end{tabular}
  \\
  Combining these uncertainties in quadrature would give 0.9\%. Our overall
  estimate of 1.5\% is somewhat more conservative.

\item 
  In total relative uncertainties of 6.4\% and 8.5\% are obtained for
  $\overline{\mathcal{B}}_{s\mu}$ and $\overline{\mathcal{B}}_{d\mu}$,
  respectively. 
\end{itemize}

\section{\label{sec::bsg}$\bar{B} \to X_s \gamma$}

\begin{figure}
  \begin{center}
    \includegraphics[width=.8\textwidth]{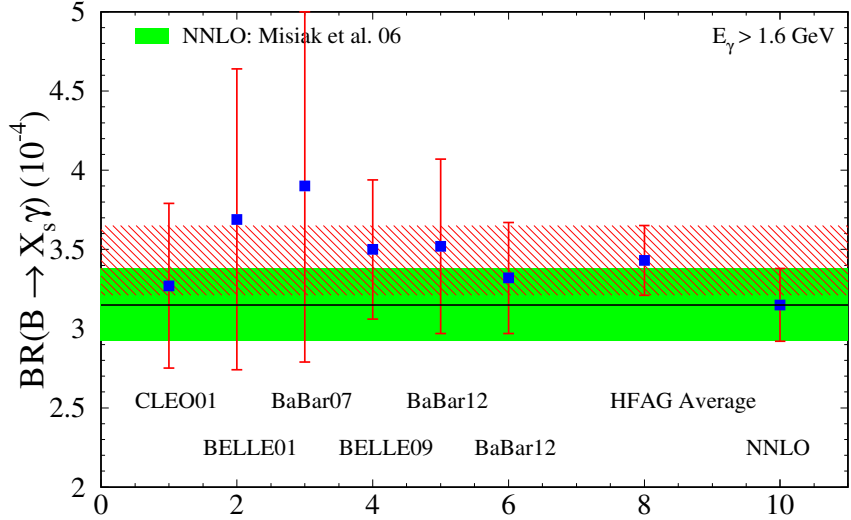}
    \caption{\label{fig::bsg}Compilation of experimental results for
      $\bar{B} \to X_s \gamma$ with a cut on the photon of $E_{\gamma} > 1.6$~GeV,
      and comparison to SM theory prediction
      from Ref.~\cite{Misiak:2006zs,Misiak:2006ab}.}
  \end{center}
\end{figure}

There are a number of experiments which have measured the branching ratio
$\overline{\mathcal B}(\bar{B} \to X_s \gamma)$ with high accuracy. A
compilation of results from CLEO~\cite{Chen:2001fja},
BaBar~\cite{Aubert:2007my,Lees:2012wg,Lees:2012ym} and
BELLE~\cite{Abe:2001hk,Limosani:2009qg} is shown in Fig.~\ref{fig::bsg} where
the hatched (red) uncertainty band corresponds to a combination performed by
the Heavy Flavour Averaging Group (HFAG)~\cite{HFAG}. The experimental result
is compared to the theory prediction of
Refs.~\cite{Misiak:2006zs,Misiak:2006ab} (solid, green band).  One observes a
significant overlap of the two bands which indicates a good agreement of the
experimental number with the SM prediction. Both uncertainty bands
amount to approximately 7\%. Once Belle~II starts data taking it is expected
that the experimental uncertainty will shrink, which calls for an improvement on the
theory side.

The theory uncertainty band in Fig.~\ref{fig::bsg} receives 
contributions from unknown higher order (3\%), input parameters (3\%),
``$m_c$-interpolation'' (3\%), and non-perturbative effects (5\%).
Near-future improvements can be expected from improved measurements of the
input parameters and new calculations of the charm-quark contributions to the
four-quark operators $Q_1$ and $Q_2$, which might improve the
``$m_c$-interpolation'' uncertainty. An analysis containing an improved
prediction is in preparation~\cite{CFHMSS}.

\end{document}